\documentclass{article}
\usepackage{ifthen}

\def \version {_astro-ph}

\ifthenelse{\equal{\version}{_apj}}
{
\usepackage{aasms4}

\newcommand{\href}[2]{{\tt #2}}

\def \figwidth {0.6 \linewidth}
}
{
\usepackage{emulateapj}
\usepackage{apjfonts}
\usepackage{hyperref}

\def \figwidth {\linewidth}
}

\ifthenelse{\equal{\version}{_working}}
{
}
{

}

\usepackage{astrobib}

\usepackage{epsfig}
\def \beginfig          {\begin{figure}}
\def \endfig            {\end{figure}}

\def \begineq           {\begin{equation}}
\def \endeq             {\end{equation}}

\def\gtorder{\mathrel{\raise.3ex\hbox{$>$}\mkern-14mu
             \lower0.6ex\hbox{$\sim$}}}
\def\ltorder{\mathrel{\raise.3ex\hbox{$<$}\mkern-14mu
             \lower0.6ex\hbox{$\sim$}}}

\def \lsim {\ltorder}

\def \hide#1{}

\def\br {{\bf r}}

\def\bx {{\bf x}}

\def\bgamma {\mbox{\boldmath $\gamma$}}

\def\btheta {\mbox{\boldmath $\theta$}}

\def\araa{{ARA\&A}}             
\def\apj{{ApJ}}                 
\def\aap{{A\&A}}                
\def\mnras{{MNRAS}}             
\def\pasp{{PASP}}               

\def \I {{\rm I}}
\def \V {{\rm V}}

\makeatletter
\newenvironment{inlinetable}{%
\def\@captype{table}%
\noindent\begin{minipage}{0.999\linewidth}\begin{center}\footnotesize}
{\end{center}\end{minipage}\smallskip}
\newenvironment{inlinefigure}{%
\def\@captype{figure}%
\noindent\begin{minipage}{0.999\linewidth}\begin{center}}
{\end{center}\end{minipage}\smallskip}
\makeatother

\begin{document}

\title{Large-Scale Cosmic Shear Measurements}

\author{Nick Kaiser, Gillian Wilson and Gerard A.~Luppino}
\affil{Institute for Astronomy, U.~Hawaii \\
2680 Woodlawn Drive, Honolulu, Hawaii 96822 \\
{\tt kaiser@hawaii.edu}, {\tt http://www.ifa.hawaii.edu/$\sim$kaiser}}


\begin{abstract} {
We present estimates of the gravitational lensing shear variance obtained
from images taken at the CFHT using the UH8K CCD mosaic camera.
Six fields were observed for a total of 1 hour each in V and I,
resulting in catalogs containing $\sim 20,000$ galaxies per field, with
properly calibrated and optimally weighted shear estimates.  These
were averaged in cells of sizes ranging from $1'.875$ to
$30'$ to obtain estimates of the cosmic shear variance 
$\langle \overline{\bgamma}^2\rangle$, with
uncertainty estimated from the scatter among the estimates for the 6 fields.
Our most reliable estimator for cosmic shear is provided by the 
cross-correlation of the shear measured in the two passbands.
At scales $\lsim 10'$ the results are in good agreement with
those of \citeN{vwme+00}, \citeN{bre00} and \citeN{wtk+00} and with
currently fashionable cosmological models.  
At larger scales the shear variance falls below 
the theoretical predictions, and on the largest scales we find 
a null detection of shear variance averaged in $30'$ cells of
$\langle \overline{\bgamma}^2 \rangle = (0.28 \pm 1.84) \times 10^{-5}$.
}\end{abstract}


\section{Introduction}

Weak lensing provides a potentially powerful probe of mass fluctuations in
the Universe (\citeNP{gunn67}; \citeNP{mellier99} and references therein).  
Three independent groups have recently presented
estimates of the shear variance from deep `blank-field' CCD imaging
surveys. 
\citeNP{vwme+00} (hereafter vWME+) measured the shear variance in circular cells
of radii ranging from $0'.7$ to $3'.5$;
\citeNP{bre00} (hereafter BRE) measured the shear variance in square cells of
side $8'.0$ and \citeNP{wtk+00} (hereafter WTK+) have provided estimates of
the shear-shear correlation function at separations $3'.25$, $8'.5$ and $22'.0$.
Here we present shear variance measurements from $\simeq 1.5$ square degrees
of deep photometry obtained as part of our ongoing weak lensing
survey.  We find results which are broadly in good agreement with the
recently published estimates.

\section{The Data}

The data were taken at the 3.6m CFHT telescope using the 
$8192 \times 8192$ pixel UH8K camera at prime focus.
The camera delivers a field size of $0^\circ.5$ with $0''.207$ pixels.
Our survey strategy has been to target blank fields in six widely separated
areas for ease of scheduling, and in each area we plan to make $6$ or so
pointings scattered over a region of extent $\sim 3^\circ$.
In January 1999 the UH8K was replaced by the CFH12K camera. 
By that time we had completed
1 hour integrations in both V and I for two pointings in each of three areas.
The field names, centers and also the estimated seeing 
are given in table \ref{tab:fields}. 
One of the 8 devices in the mosaic (lying in the NW corner of our images)
has very poor charge transfer efficiency and the data from this device
were discarded.  After further masking of regions around
bright stars the total useful solid angle per field is $\simeq 0.16$
square degrees.

\begin{table*}[htbp!]
\begin{center}
\caption{Field centers and seeing}
\begin{tabular}{cccccccc}
\hline
\hline
area & pointing	& RA (J2000)	& DEC (J2000)	& $l$		& $b$	& FWHM(I)	& FWHM(V)	\\
\hline
1650	& 1	& 16:51:49.0	& 34:55:2.0 	& $57.37$	& $38.67$	& $0''.82$	& $0''.85$	\\
	& 3	& 16:56:0.0	& 35:45:0.0 	& $58.58$	& $37.95$	& $0''.85$	& $0''.72$	\\
Groth	& 1	& 14:16:46.0	& 52:30:12.0	& $96.60$	& $60.04$	& $0''.80$	& $0''.93$	\\
	& 3	& 14:9:0.0	& 51:30:0.0 	& $97.19$	& $61.57$	& $0''.70$	& $0''.85$	\\
Lockman	& 1	& 10:52:43.0	& 57:28:48.0 	& $149.28$	& $53.15$	& $0''.83$	& $0''.85$	\\
	& 2	& 10:56:43.0	& 58:28:48.0 	& $147.47$	& $52.83$ 	& $0''.84$	& $0''.86$ 	\\	
\hline
\end{tabular}
\label{tab:fields}
\end{center}
\end{table*}

\section{Data Reduction}

The data were reduced much as for our MS0302 supercluster observations \cite{kwld99}.
After flat fielding, an object finder was applied to each image, 
and a set of bright but non-saturated stars extracted for registration
purposes.  The positions of these stars, along with celestial coordinates from the
USNOA \cite{usnoa} catalog, were used to find a mapping from image pixel coordinates to
orthographic sky coordinates.  

Images of the stars were analyzed to generate
a model for the point spread function
(PSF) $g(\bx; \br)$, this being the 2-dimensional profile of a star
with centroid at $\br$ measured in coordinates $\bx$ being measured relative to the
centroid.  The model is a sum of 2-D image valued modes $g_i(\bx)$:
$g(\bx; \br) = \sum_i c_i(\br) g_i(\bx)$
with coefficients $c_i(\br)$ which are low order polynomials in star position.
We found a 1st order model to be adequate to describe the variation of
the PSF with position on the chip (though see below). 

The astrometric solutions, the PSF models, and also standard
star observations were first used to 
create a set of photometrically calibrated `homogenized' images which were
degraded to have a common identical PSF.  These images were 
compared in order to identify cosmic rays
and other transient events.
Next, for each raw image we generated a re-circularized image
by convolving with a 90-degree rotated version of the PSF model.  These, as well as the
raw images, were then warped to sky coordinates, with previously identified cosmic rays
being removed, and the stacks of images combined to provide a quilt of overlapping
images.  This procedure results in three images: a median of the raw images
(in which the PSF is generally
non-circular), a median of the re-circularized images, 
and also a sky-noise image.
The final summed images were sampled with $0''.15$ pixel size.

The final object catalogs were obtained by applying {\tt hfindpeaks}
to the median averaged raw images, this program having been modified in
order to allow properly for the non-trivial noise correlations
in the final images.
After applying aperture photometry analysis, shapes of the objects were
measured as described
in \citeN{kaiser00}.  The essential result of this is a polarization vector
$q_\alpha = M_{\alpha l m} \int d^2 x\; w(x) x_l x_m f(\bx)$ 
formed as a combination of weighted second moments of the re-circularized image,
and a polarizability tensor which describes the response of the
polarization to gravitational shear.
The weight function $w(x)$ was taken to be a Gaussian ball of 2 pixels in scale length.
For convenience, the quantities actually generated were the normalized polarization
$\hat q_\alpha = q_\alpha / \sqrt{q_\beta q_\beta}$ and a
polarizability $Q$ defined such that for galaxies with this shape and size,
the expectation value of $\hat q_\alpha$ is $\langle \hat q_\alpha \rangle
= Q \gamma_\alpha$.  The $q_\alpha$ values were corrected for
artificial shear introduced in the image warping by
slight errors in our astrometric solution.

Finally a selection on significance of $4 \le \nu \le 100$ was applied to
select a `faint galaxy' catalog.  The corresponding magnitude limits are
somewhat fuzzy since significance depends on both size and magnitude.
The counts (number per magnitude interval) turn over at about $m_\I \simeq 24$ and 
$m_\V \simeq 25$.  Numbers of objects in the final catalogs are shown in
table \ref{tab:galcats}. The density of objects on the sky is very similar to
that in the images obtained by vWME+, WTK+, and slightly higher than the density
of objects in the BRE sample, so our shear
variance estimates should be more or less directly comparable.

Given a single galaxy, a fair (but very noisy) estimate of the shear 
is $\gamma_\alpha = \hat q_\alpha / Q$.  However, 
since the normalized polarization response $Q$ varies with shape and
size of the object (small objects having very little response for example),
to measure the mean shear in a region containing $N$
galaxies one should weight the individual estimates by $Q^2$ so the
optimal mean shear estimate is 
$\overline{\gamma}_\alpha = \sum Q \hat q_\alpha / \sum Q^2 $.
This assumes that one has little prior knowledge of the
of galaxy redshifts.  
If the averaging region contains a large number of galaxies, as
is the case for the cells considered here, one can replace 
$\sum  Q^2$ by $N \langle Q^2 \rangle$ 
where the $\langle Q^2 \rangle$ is an average over all of the
galaxies in the catalog.
The optimal mean shear is then the average of 
weighted shear values for the individual galaxies: 
$\gamma_G = \omega_G \hat q_G / Q_G$, with normalized weight
$\omega_G \equiv Q_G^2 / \langle Q^2 \rangle$.
The quantity $\langle Q^2\rangle$ 
is a useful measure of the image quality.  It is equal to
the inverse shear variance per galaxy, so, for instance, the statistical
uncertainty in the mean shear $\overline{\gamma}$ 
measured from a sample of $N$ galaxies is
$\sigma^2_{\overline{\gamma}} = 1 / (N \langle Q^2 \rangle)$.  The values
of $\langle Q^2 \rangle$ are also given in table \ref{tab:galcats}.
 
Preliminary results of this analysis \cite{wkl99} in the form of estimates
of the net shear for each of the six fields and for catalogs
generated from the I and V images separately gave shear values typically
of about 1\%, but with a few larger values.  These, however, showed little
correlation between the two passbands, suggesting that the results were
contaminated by some systematic error.  Examination of the
re-circularized images of the stars in the fields with seemingly spurious
shear values revealed at least a major part of the problem.  
The mean stellar polarization
was found to vary systematically with magnitude.  This is to be expected for
very bright stars where the pixels saturate and charge begins to bleed along
the slow direction of the CCD.  The effect found here had the same signature
(a trend for $q_1$ to become negative for bright objects) but appeared
at a low level and, unexpectedly, for stars much fainter than the saturation
limit.  The result was that in some cases our PSF model fitting procedure,
which weighted stars according to brightness, did not correctly recircularize
the faint stars as it should (since we require the PSF appropriate for
faint galaxies). 
The effect seemed to be variable, and also
tended to be associated with particular chips.
As a simple fix, 
we fit the residual polarizations for the
faintest stars ($I > 18$, $V > 21$) to a 4th order spatial 
polynomial and then used a
smear polarizability analogous to that defined by 
\citeN{ksb95} to correct the galaxy $q_\alpha$ values.
This reduced the spurious shear values considerably.  

\begin{inlinetable}
\caption{Galaxy catalogs}
\begin{tabular}{ccccc}
\hline
\hline
field	& $N_\I$	&	$\langle Q^2 \rangle_\I$	& $N_\V$	&	$\langle Q^2 \rangle_\V$ \\
\hline
  1650 1 &  21569 &      1.527  	& 15403 &      1.234 \\
  1650 3 &  18187 &      1.785 		& 16518 &      1.147 \\
 Groth 1 &  27293 &      1.440 	& 16391 &      1.543 \\
 Groth 3 &  19162 &      1.254 	& 15876 &      1.775 \\
Lockman 1 &  20726 &      1.855 	& 20358 &      1.352 \\
Lockman 2 &  20017 &      1.630 	& 17779 &      1.417 \\
\hline
\end{tabular}
\label{tab:galcats}
\end{inlinetable}

\section{Cosmic Shear Variance}

We have chosen to focus here on a single simple statistic: the variance of the
shear averaged in cells of various sizes.   This is the statistic used by
vWME+ and BRE and is simply related to the shear covariance
function presented by WTK+. The cell averaged shear variance is also
simply computable from the spectrum of mass fluctuations (e.g.~\citeNP{k92b}), so this
provides a useful link between observation and theory.
Now each weighted shear estimate $\gamma_G$ consists of a random intrinsic
component $\gamma_{G{\rm int}}$ 
and a `cosmic' component proportional to the integral of the tidal field
along the line of sight.  Modeling the cosmic shear 
as the sum over a set of statistically independent screens, we have
\begineq
\gamma_G = \gamma_{G{\rm int}} + \omega_G
\sum_S \gamma_S(\btheta_G) \beta(z_G, z_S)
\endeq
where $\gamma_S(\btheta)$ is the shear field for 
the $S$th screen and for fictitious sources at
infinite distance, and $\beta \equiv {\rm max}(0, 1 - D_{SG} / D_{OG})$ is 
the usual ratio of angular diameter distances.

This model allows one to compute the variance of the shear averaged
over galaxies falling in a cell on the sky.  We will also be interested
in the co-variance of shear measured in different passbands.
Consider the mean shear 
$\overline{\gamma}_P = (1 / N_P) \sum \gamma_P$ for a specific cell and
for galaxies found in two passbands $P = A, B$.
Averaging over realizations of random intrinsic shear values,
and also averaging over an ensemble of realizations of cosmic shear
screens, yields
\begineq
\label{eq:shearvar0}
\langle \overline{\bgamma}_A \cdot \overline{\bgamma}_B \rangle
= {N_{AB} \over N_A N_B} \langle \bgamma_A \cdot \bgamma_B \rangle +
\sum\limits_S 
\langle \omega_A \beta_{AS} \rangle
\langle \omega_B \beta_{BS} \rangle
\langle \overline{\gamma}_S^2 \rangle
\endeq
where $N_{AB}$ is the number of objects in the cell which were detected in both
passbands.  This formula is also valid when $A$ and $B$ are the same.
The expectation value of the dot product of the cell averaged
shear is therefore equal to a noise term plus a 
cosmic term which is a sum of the cell-averaged shear variances for the screens.
Interestingly, the noise term involves the total shear
variance $\langle \bgamma_A \cdot \bgamma_B \rangle$, containing both
intrinsic
and cosmic contributions, which is convenient since this is the quantity
that one can actually measure.
In obtaining (\ref{eq:shearvar0}) we assumed that the faint galaxies
are randomly distributed on the sky, this being motivated by the
fact that the angular correlation function is very small,
with $w(\theta) \lsim 10^{-2}$
on all relevant scales. 

To implement this, for each field we averaged the shear in a 
grid of contiguous square cells of side $L$, and 
those cells in the lower quartile of occupation number were discarded.  
To obtain an estimate of the 
cosmic shear variance we then computed for each field
\begineq
\label{eq:covariance}
\langle \overline{\bgamma}^2 \rangle_{AB}  = 
{1 \over n_{\rm cells}} \sum\limits_{\rm cells} \left[
\overline{\bgamma}_A \cdot \overline{\bgamma}_{B} - 
N_{AB} \langle \bgamma_A \cdot \bgamma_B \rangle/ (N_A N_B) \right].
\endeq
The shear covariance functions are
$\langle {\bgamma}_\I \cdot {\bgamma}_\I \rangle = 1/ \langle Q^2 \rangle_\I$ and
$\langle {\bgamma}_\V \cdot {\bgamma}_\V \rangle = 1/ \langle Q^2 \rangle_\V$
whereas $\langle {\bgamma}_\I \cdot {\bgamma}_\V \rangle$ was 
estimated by correlating the shears for objects which were
detected in both the I and V catalogs.
The shear variances estimated from the separate fields
were then averaged together to obtain a final cosmic shear variance with
uncertainty estimated from the scatter of the field estimates about the mean.

The diagonal components of
$\langle \overline{\bgamma}^2 \rangle_{AB}$ provide
estimates for the shear variance for the respective passbands, 
with strength roughly proportional 
to the square of the mean distance to the galaxies
(assuming a spectrum of mass fluctuations with index $n \simeq -1$), and
the off-diagonal components should lie somewhere in between. 
The results are shown in figure \ref{fig:shearvar1} and in table \ref{tab:vardata}
and deviate somewhat from
this expectation: the I-V cross correlation 
lies systematically below both the I- and V-band
shear variance estimates.
These results are robust to changes in the
order of the polynomial in the stellar polarization model, and they
are not caused by a few discrepant cells. 

The difference between the I and V band shear variance may be due to
differences in the redshift distributions, some evidence
for which was found by \citeN{lk97} in their study of MS1054.  However, 
we typically find about 60\% of the galaxies are detected in both passbands,
so this requires fairly high redshifts for the blue galaxies.
For example, assume that the I-band sample has a redshift distribution
like that measured by Cowie (personal communication) in the range $23 < \I_{\rm AB} < 24.0$, but that
the V band sample contains an additional 40\% population of higher redshift galaxies.
If we place these at redshift 3, we find that the V-band shear variance is
about a factor 2 higher than the I-band, much as seen in figure \ref{fig:shearvar1}. 
However the I-V cross-correlation is then predicted 
to be about 30\% higher than the I-band variance, which is not seen.

The simplest
interpretation of these results is that the shear inferred from the
I- and V-band data separately has been inflated by 
residual systematic errors of some kind.
The level of these errors
is on the order of 2 percent rms shear on scales of a few arc-minutes, falling to
somewhat below the 1 percent level on $30'$ scales.
If so, the most reliable estimate of the
cosmic shear variance 
is provided by the I-V cross-correlation since systematic errors which
are uncorrelated between the passbands will cancel out.  Of course there is no
guarantee that the cross-correlation is not affected by some source of error
which is common to both passbands, a specific example of which is 
artificial shear arising from intrinsic
correlation of galaxy shapes in clusters etc.~due to tidal effects.

\begin{inlinefigure}
\centering\epsfig{file=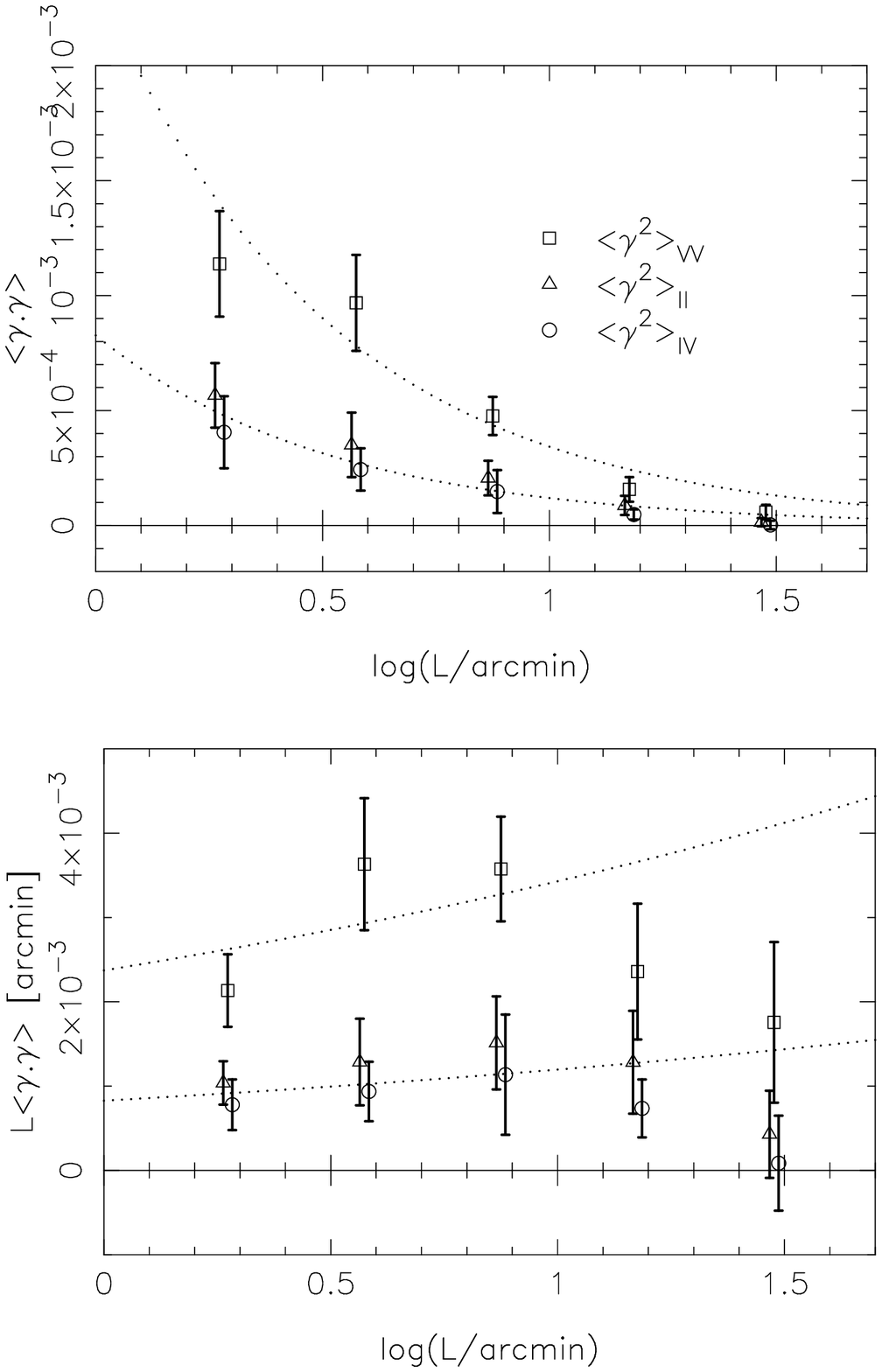,width=\figwidth}
\caption[Shear covariance matrix.]
{Estimates of the cosmic shear variance as a function of averaging scale $L$.
Lower panel shows the same data multiplied by $L$ to show more clearly the
large angular scale results.  The dotted lines are predictions for 
a currently fashionable cosmological model --- cluster normalised;
$\Omega_{\rm m} = 0.3$; $\Omega_\Lambda = 0.7$; $\Gamma = 0.25$ --- from
\citeN{js97} for sources with an effective redshift (i.e.~equivalent
single screen redshift) of $z_{\rm eff} = 1.0$ and $z_{\rm eff} = 2.0$.
The points have been displaced slightly laterally for clarity.}
\label{fig:shearvar1}
\end{inlinefigure}

\begin{inlinetable}
\caption{Shear variance}
\begin{center}
\begin{tabular}{cccc}
\hline
\hline
$L$/arcmin & $10^4 \times \langle \overline{\gamma}^2 \rangle_{\I\I}$ &
 $10^4 \times\langle \overline{\gamma}^2 \rangle_{\V\V}$ &
 $10^4 \times\langle \overline{\gamma}^2 \rangle_{\I\V}$ 	\\
\hline
$            30$ & $      0.15 \pm       0.18$ & $      0.59 \pm       0.32$ & $      0.03 \pm       0.18$  \\
$            15$ & $      0.87 \pm       0.42$ & $      1.57 \pm       0.54$ & $      0.48 \pm       0.22$  \\
$           7.5$ & $      2.06 \pm       0.75$ & $      4.77 \pm       0.83$ & $      1.48 \pm       0.93$  \\
$          3.75$ & $      3.51 \pm       1.40$ & $      9.69 \pm       2.09$ & $      2.44 \pm       0.92$  \\
$         1.875$ & $      5.66 \pm       1.40$ & $     11.38 \pm       2.29$ & $      4.06 \pm       1.57$  \\
\hline
\end{tabular}
\label{tab:vardata}
\end{center}
\end{inlinetable}

\begin{inlinefigure}
\centering\epsfig{file=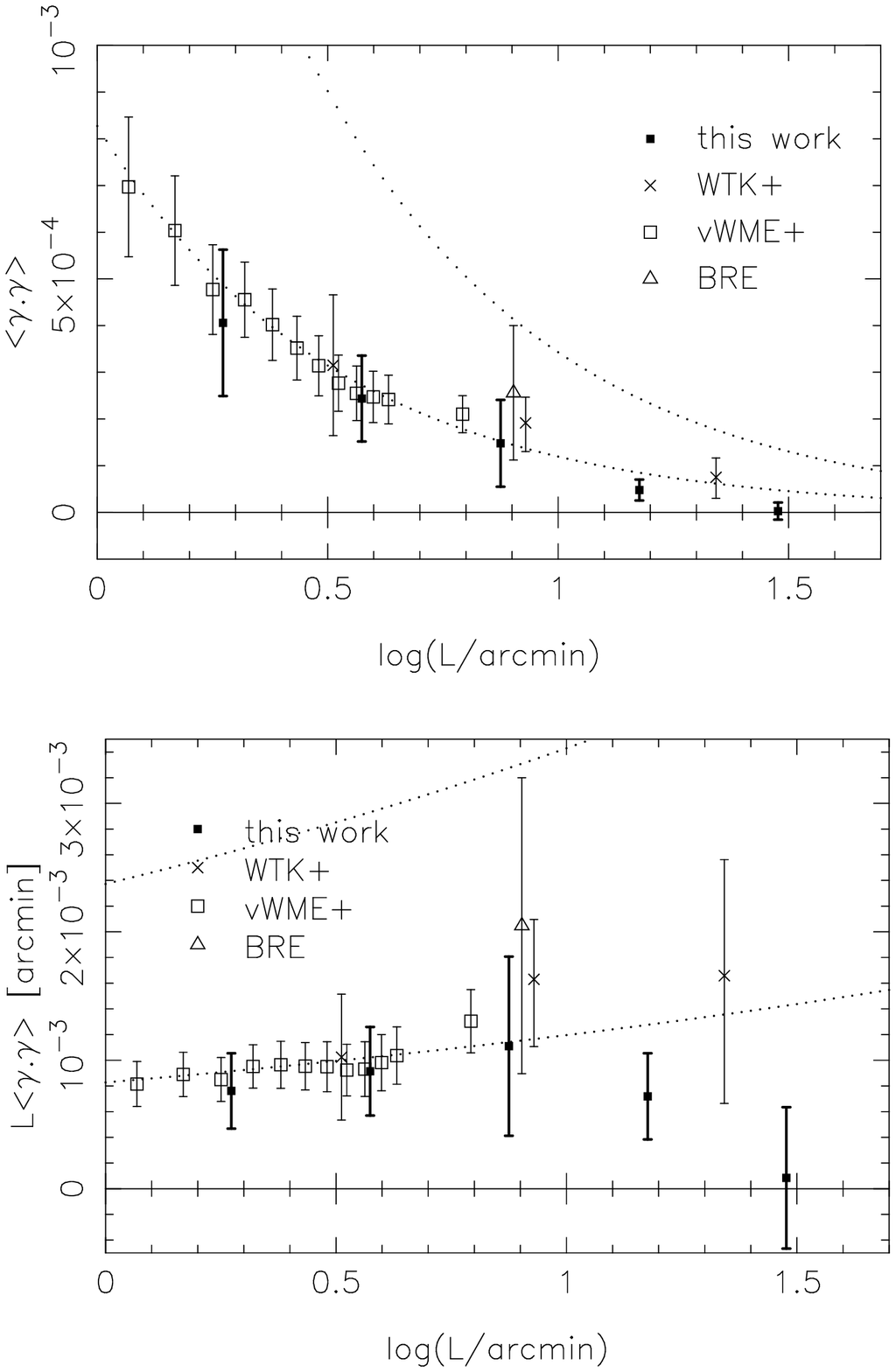,width=\figwidth}
\caption[Shear variance vs scale]
{Our estimates of the cosmic shear variance from the I-V cross-correlator
are shown as the heavy points. Also shown are results from
vWME+, BRE, WTK+.  The error bars
on the vWME+ estimates are statistical only. All others are
total error including cosmic variance.
The lower panel again shows the same data multiplied by $L$ to show more clearly the
large angular scale results. 
The dotted lines are the \citeN{js97} predictions as in figure 
\ref{fig:shearvar1}.}
\label{fig:shearvar2}
\end{inlinefigure}

The I-V shear variance estimator is shown with an expanded vertical scale in figure
\ref{fig:shearvar2}.  Also shown are the recently announced
results.  The BRE result is shown as presented in their paper and with total
error estimate including cosmic variance. The vWME+ circular cell average shear
are plotted against $L = \sqrt \pi \theta$.  The vWME+ error bars are statistical
only.
WTK+ presented estimates of the ellipticity
correlation function $C_1(\theta) = \langle \epsilon_1(0) \epsilon_1(\theta) \rangle$. 
We have converted their $C_1(\theta)$ to an equivalent
shear variance using formulae from \citeN{k92b} with $\epsilon = 2 \gamma$
and  assuming a spectral index $n = -1$.  The lower panel shows the variance
multiplied by averaging box size $L$.  For a $n = -1$ spectrum, corresponding to a 
mass auto-correlation function $\xi(r) \propto 1 / r^2$, this 
quantity should independent of scale.

At small scales $\lsim 10$ arcmin there seems to be remarkably good agreement 
between the independent estimates.  Note that the measurements were made using
three separate observing facilities.
At $L = 3'.75$ we find 
$\langle \overline{\bgamma}^2 \rangle \simeq 2.5 \simeq 10^{-4}$.
This about a factor 4-5 lower than the 
prediction for a light-traces mass $\Omega_m = 1$ cosmology, 
and an effective redshift for the background
galaxies $z_{\rm eff} = 1$ \cite{k92b,js97}.

At larger scales the shear variance we find falls below that
of WTK+.  Their largest scale estimates appear to conflict
with our null result at about the 2-sigma level.
Our large-angle results are also smaller than the
$\Omega_{\rm m} = 0.3$, $\Omega_\Lambda = 0.7$ theoretical model predictions.

\section{Discussion}

For an effective background galaxy redshift of $z_{\rm eff} \simeq 1.0$
these measurements probe mass fluctuations in a shell
peaked at $z \simeq 0.4$.
At this redshift the $30'$ field size corresponds 
to a comoving distance of about $6 h^{-1}$Mpc,
so the cell variances presented here probe scales in the range $0.4-6 h^{-1}$Mpc.
On the smaller end of this scale we find very good agreement with recently
announced estimates from other groups, and also with canonical cosmological theory
predictions.  It is hard to definitively rule out the possibility that the
small angle measurements are inflated by systematic errors, but one can safely
rule out theories such as light-traces mass high density models
which predict shear variance a factor $\sim 5$ higher than our results.

On larger scales our measurements are extremely precise, yet we find only
a null detection for our largest cells.  These results show that on large scales
the rms shear is at most a fraction of a percent.
The apparent discrepancy between these results and the theoretical predictions
is quite interesting, and suggests a steepening of the mass correlation
function at scales $\sim 1-2 h^{-1}$Mpc.  More data are needed however
to definitively confirm this.

\section{Acknowledgements}

The results here were extracted from data taken at the Canada France Hawaii Telescope.
The analysis was supported by NSF grants AST95-00515, AST99-70805.
GW gratefully acknowledges financial support from the
estate of Beatrice Watson Parrent and from Mr.~\& Mrs.~Frank 
W.~Hustace, Jr. We thank Peter Schneider and Gary Bernstein for helpful
suggestions.

\ifthenelse{\equal{\version}{_working}}
{
\bibliographystyle{apj}
\bibliography{astro,clusters,weaklensing,kaiser_ref,kaiser_nonref}
}
{

}

\end{document}